\documentclass[twocolumn,aps,showpacs,preprintnumbers,nofootinbib,
superscriiptaddress]{revtex4}
\usepackage{graphicx}
\usepackage{dcolumn}
\usepackage{bm}
\begin{document}

\title{The coherent synchrotron radiation influence on the storage ring
longitudinal beam dynamics}

\author{E.G. Bessonov, R.M. Feshchenko, \\
{\it Lebedev Physical Institute RAS, Moscow, Russia} \\
V.I. Shvedunov         \\
{\it Moscow state University}}


                       \begin{abstract}
We investigate influence on the storage ring beam dynamics of the coherent
Synchrotron Radiation (SR) self fields produced by an electron bunch. We
show that the maximum energy gain in the RF cavity must far exceed the
energy loss of electrons due to the coherent SR.
\end{abstract}


\maketitle

\section{Introduction}

The energy $\varepsilon $ of a particle in storage rings oscillates in the
vicinity of the equilibrium energy$\varepsilon _s $. The difference between
equilibrium and nonequilibrium energies is proportional to the derivative of
the particle's phase $d\varphi / dt = h(\omega _s - \omega _r )$:

\begin{equation}
\label{eq1}
\Delta \varepsilon = \varepsilon - \varepsilon _s = \frac{\varepsilon _s
}{hK\omega _s }\frac{d\varphi }{dt},
\end{equation}
\noindent
where $K = - \partial \ln \omega _r / \partial \ln \varepsilon = (\alpha
\gamma _s^2 - 1) / (\gamma _s^2 - 1)$ is self phasing coefficient; $\alpha $,
the momentum compaction factor; $\varphi = \int {\omega _r (t)dt} $,
the particle's phase; $\gamma = \varepsilon / mc^2$, the relative energy;
$\omega _r = 2\pi f$; $f$, the revolution frequency of a particle in the
storage ring. Equilibrium values have lower index $s$ [1-3]. The radio
frequency (RF) voltage in the cavity's gap is varying as $V = V_{rf} \cos
\omega _{rf} t$, where $\omega _{rf} $ is the radio frequency; $h$, the
subharmonic number of radio frequency.

Balance of energy gained by an electron during the period of a single
revolution $T = 1 / f = C / c = 2\pi R(1 + \mu ) / c$ in the RF cavity and
lost due to synchrotron radiation and Thomson scattering defines an equation
for electron phase oscillations in the storage ring:

\begin{equation}
\label{eq2}
\frac{d\varepsilon }{dt} = \frac{eV_{rf} \cos \varphi }{T} - < P^{rad} > ,
\end{equation}
\noindent
where $ < P^{rad} > = d\varepsilon ^{rad} / dt$ is the power of radiation
losses averaged over the length of the orbit; С, the length of the orbit; R,
the curvature radius of the particle orbit in bending magnets; $\mu =
\sum\nolimits_i {l_i / 2\pi R} $, the ratio of the sum of straight intervals
$l_i $ in the storage ring to the path length in the bending magnets. The
synchronous phase $\varphi _s $is defined as $d\varepsilon _s / dt = 0$ or
$eV_{rf} \cos \varphi _s = \left\langle {P_s^{rad} } \right\rangle T$.

The spontaneous coherent SR doesn't depend on the particle energy but
depends on the particle position in the longitudinal direction, the shape of
the beam and on the number of particles. For the Gauss longitudinal
distribution one can obtain:

\[
\left\langle {P_{coh}^{rad} \left( \varphi \right)} \right\rangle = -
\frac{3^{1 / 6}\Gamma ^2\left( {2 / 3} \right)Ne^2c}{2^{1 / 3}\pi R^{2 /
3}\sigma _s^{4 / 3} \left( {1 + \mu } \right)}*\]

$$
\exp \left[ { -
\frac{1}{2}\left( {\frac{R\left( {\varphi - \varphi _s } \right)\left( {1 +
\mu } \right)}{h\sigma _s }} \right)^2} \right]*
$$

$$
[ 1 - \frac{2^{1 / 6}\sqrt \pi }{3\sqrt 3 \Gamma \left( {2 / 3}
\right)}\frac{R\left( {\varphi - \varphi _s } \right)\left( {1 + \mu }
\right)}{h\sigma _s } -
$$

\begin{equation}
\label{eq3}
\frac{1}{6}\left( {\frac{R\left( {\varphi - \varphi
_s } \right)\left( {1 + \mu } \right)}{h\sigma _s }} \right)^2 + \ldots
],
\end{equation}

It is supposed here that the phase in the center of the bunch is equal to
synchronous phase $\varphi _s $, $\sigma _s $ is the bunch mean square
length and $\Gamma (2 / 3) = 1.35$ [4].

If the laser beam is homogeneous and its transversal dimensions far exceed
ones of the electron beam, the powers of Thomson scattering radiation and
spontaneous incoherent SR obey the simple power dependence as functions of
energy$ < P_{noncoh}^{rad} > = < P_{s,noncoh}^{rad} > (\varepsilon /
\varepsilon _s )^{k_i }$. The difference between radiated power of
synchronous and nonsynchronous particles is

     $$ < P_{noncoh}^{rad} > - < P_{s,noncoh}^{rad} > = \frac{d <
     P_{s,noncoh}^{rad} > }{d\varepsilon }\Delta \varepsilon =$$

\begin{equation}
\label{eq4}
 k_i <
P_{s,noncoh}^{rad} > \frac{\Delta \varepsilon }{\varepsilon _s },
\end{equation}

\noindent
where $k_i = 2$ for the Thompson backscattering, $k_{i} =1$ for the Raleigh
backscattering by ions and k$_{i}$ = 1$\div $ 1.5 for the SR.

Subtracting the power balance equation for synchronous particles from the
equation for nonsynchronous one (\ref{eq2}) and taking into account (\ref{eq1}), (\ref{eq4})
we obtain equation for phase oscillations in the storage ring:

\begin{equation}
\label{eq5}
\frac{d^2\varphi }{dt^2} + \frac{k_i < P_{noncoh}^{rad} > }{\varepsilon _s
}\frac{d\varphi }{dt}
 - \frac{he\omega _s^2 K}{2\pi \varepsilon _s }[V(\varphi ) - V(\varphi _s
)] = 0,
\end{equation}

\noindent
where $V\left( \varphi \right) = V_{rf} \cos \varphi - {2\pi R\left( {1
+ \mu } \right)}/{c}\left\langle {P_{coh}^{rad} \left( \varphi \right)}
\right\rangle $. The synchronous phase is determined by the equation $U
(\phi _s) =0$.

Incoherent synchrotron radiation and Thompson scattering cause slow damping
of phase oscillations (the damping time far exceeds the period of
oscillations) and can be neglected in the first approximation, so equation
(\ref{eq5}) can be rewritten as:

\begin{equation}
\label{eq6}
\frac{1}{2}\frac{d}{dt}(\frac{d\varphi }{dt})^2
 - \frac{he\omega _s^2 K}{2\pi \varepsilon _s }[V(\varphi ) - V(\varphi _s
)]\frac{d\varphi }{dt} = 0.
\end{equation}

The first integral, determining particle phase trajectories behavior is

\begin{equation}
\label{eq7}
\frac{d\varphi }{dt}
 = \sqrt {\frac{he\omega _s^2 K}{\pi \varepsilon _s }\int {[V(\varphi ) -
V(\varphi _s )]d\varphi } } .
\end{equation}

The coherent synchrotron radiation force acts similar to the radio frequency
accelerating field. The autophasing force of the storage ring is defined by
the derivative $dV(\varphi ) / d\varphi $. Thus, in accordance with (\ref{eq5}),
the reaction of the coherent SR makes this force weaker. This weakening
reaches maximum when the phase equals $\varphi = \varphi _s + h\sigma
_s / R\left( {1 + \mu } \right)$. Therefore the stability of the
electron beam requires that the derivative $dV(\varphi ) / d\varphi $
is negative in the interval $\vert \varphi - \varphi _s \vert < h\sigma
_s / R(1 + \mu )$. This phase range corresponds to the stable
oscillations of the majority of particles with amplitudes $A \simeq
\sigma _s $. Using the formulas for the power of the coherent SR (3)
for a beam with Gauss longitudinal distribution of particles one can
find:

\begin{equation}
\label{eq8}
V_{rf} > V_{rf,c} = \frac{2\pi R^2\left( {1 + \mu } \right)^2P_{coh}^{rad}
\left( {\varphi _s } \right)}{\sqrt {e_n } h\sigma _s },
\end{equation}

\noindent
where $e_n \approx 2.72$ is the natural logarithm foundation. In reality the
coherent SR can be neglected if the value $V_{rf} $ is 2 $\div $ 3 times
higher than $V_{rf,c} $ and $\sin \varphi _s \approx 1$. The maximum energy
gains in the RF cavity, according to (\ref{eq8}), must far exceed the energy loss
of electrons due to the coherent SR.

If the value $P_{coh}^{rad} $ is neglected, the equation (\ref{eq5}) is
transformed into the equation of small amplitude phase oscillations:

\begin{equation}
\label{eq9}
\frac{d^2\psi }{dt^2} + \frac{k_i P_{noncoh}^{rad} }{\varepsilon _s
}\frac{d\psi }{dt} + \Omega ^2\psi = 0,
\end{equation}

\noindent
where $\Psi = \varphi - \varphi _s < < 1$ and $\Omega = \omega _s \sqrt
{qhKV_{rf} \sin \varphi _s / 2\pi \varepsilon _s } $.

The equation (\ref{eq9}) has solutions that can be expressed as $\psi = \psi _m
(t)\cos \Omega ^{'}t$, where $\psi _m = \psi _{m,0} \exp ( - t / \tau
_{ph} )$ is the varying amplitude and

\begin{equation}
\label{eq10}
\tau _{ph} = \frac{\varepsilon _s }{P_{noncoh}^{rad} },
\end{equation}

\noindent
the damping time, $\Omega ^{'} = \sqrt {\Omega ^2 + \tau _s ^{ - 2}} $
, the frequency of small particle oscillations.

\begin{center}
\textbf{Example}
\end{center}

An electron storage ring has the radius R=50 cm, h=10, $\sigma _s $= 1 cm,
$\mu = 1$, $N=10^{10}$,$\sin \varphi _s \approx 1$. In this case the losses of
a synchronous particle per a revolution is $V_{coh}^{rad} (\varphi _s )$=
9.25 kev, $V_{rf} > 114$ kV. Thus for the stable storage ring operation the
RF cavity voltage should be much higher than the coherent radiation losses.
The shielding by the vacuum chamber can weaken this requirement [5]. One
should also note that the energy losses of a synchronous electron per a
revolution are approximately $2^{2 / 3}$ times greater than average losses
of electrons in the beam (see Appendix).

\begin{center}
\textbf{Appendix}
\end{center}

Suppose that a beam has small angular \textit{$\Delta \theta \sim $1/$\gamma $} and energy \textit{$\Delta \varepsilon $/$\varepsilon \sim $1/$\gamma $} spread (emittance). In
such a case electromagnetic fields emitted by different particles are
similar to each other but have a temporal shift. The Fourier images of these
fields are:${\rm {\bf E}}_{i,\omega } = {\rm {\bf E}}_{1,\omega } \exp
(i\Delta \varphi _i )$ i=1,2,3, ... N, where the phase difference between
waves emitted by the first and the i-th particles is $\Delta \varphi =
\omega (t_i^{'} - t_1^{'} ) + {\rm {\bf k}}[{\rm {\bf r}}(t_i^{'} ) - {\rm {\bf
r}}(t_1^{'} )]$. The moments of emission $t$ and detection $t{'}$ are connected
as $t = t{'} - R_0 / c - {\rm {\bf nr}} / c$, $R_0 $ is the distance
between the points of emission and detection, ${\rm {\bf k}} = \omega
\cdot {\rm {\bf n}} / c$, ${\rm {\bf n}}$ is a unit vector pointing in
the direction of emission, \textbf{r }-- the vector lying in the plane
perpendicular to the trajectory of a particle. The time difference for
ultrarelativistic particles $t_i^{'} - t_1^{'} $ is connected with the
space distance by a simple relation $c(t_i^{'} - t_1^{'} ) = z_i - z_1 $.
Therefore the Fourier image of the sum of fields of N particles ${\rm
{\bf E}}_\omega = \sum\nolimits_i {{\rm {\bf E}}_{i,\omega } } $ can be
written as (for the electrical field):

\begin{equation}
\label{eq11}
{\rm {\bf E}}_\omega = N\int\limits_{ - \infty }^\infty {\rho (z,{\rm {\bf
r}}){\rm {\bf E}}_{1,\omega } \exp [i\Delta \varphi (z,{\rm {\bf r}})]dz}
d{\rm {\bf r}},
\end{equation}

\noindent
where $\rho (z,{\rm {\bf r}})$ -- the density distribution of particles
normalized to unity.

If the transversal dimensions of the beam are small, the integration in the
equation (\ref{eq11}) over transversal coordinate \textbf{r }can be omitted:

\begin{equation}
\label{eq12}
{\rm {\bf E}}_\omega = N \cdot {\rm {\bf E}}_{1,\omega } \int\limits_{ -
\infty }^\infty {\rho (z)\exp } [i\frac{2\pi z}{\lambda }]dz.
\end{equation}

In this case the spectra-angular distribution of the emitted energy

\begin{equation}
\label{eq13}
\frac{\partial ^2\varepsilon ^{coh}}{\partial \omega \partial o} = cR_0^2
\vert {\rm {\bf E}}_{1,\omega } \vert ^2 = N^2\frac{\partial ^2\varepsilon
_1 }{\partial \omega \partial o}s(\omega ),
\end{equation}

\noindent
where $\varepsilon _1 $ is the energy of the radiation emitted by a single
particle, $s\left( \omega \right) = \left| {\int_{ - \infty }^\infty {\rho
\left( z \right)\exp \left[ {i2\pi z / \lambda } \right]dz} } \right|^2$,
the spectral radiation coherence factor, $\lambda = 2\pi c / \omega $-- the
wavelength of SR. The spectral energy distribution and the full emitted
energy can be found by integration of (\ref{eq13}) over angles

\begin{equation}
\label{eq14}
\frac{\partial \varepsilon ^{coh}}{\partial \omega } = N^2\frac{\partial
\varepsilon _1 }{\partial \omega }s(\omega )
\end{equation}

\noindent
and over frequency

\begin{equation}
\label{eq15}
\varepsilon ^{coh} = N^2\int\limits_0^\infty {\frac{\partial \varepsilon _1
(\omega )}{\partial \omega } \cdot s(\omega )} \cdot d\omega .
\end{equation}

From (\ref{eq13}) -- (\ref{eq15}) it follows that for a point-like beam $\rho (z) =
\delta (z)$ and therefore $s(\omega ) = 1$, $\varepsilon ^{coh} =
N^2\int_0^\infty {[\partial \varepsilon _1 (\omega ) / \partial \omega
]d\omega } = N^2\varepsilon _1 (\omega )$, i.e. the energy emitted by the
beam is N$^{2}$ times larger than the energy emitted by a single particle.

If the beam's motion is periodical one can introduce average radiation
power:$P^{coh} = f \cdot \varepsilon ^{coh}$, $\partial P^{coh} / \partial
\omega = f \cdot \partial \varepsilon ^{coh} / \partial \omega $, где $f = v
/ C$ -- the revolution frequency, $\nu \approx c$-- the particle's velocity
and $C$ is the perimeter of the orbit.

The values $\partial \varepsilon _1 / \partial \omega $ and $\partial P_1 /
\partial \omega = f \cdot \partial \varepsilon _1 / \partial \omega $ are
known. In particular, the spectral power of radiation is

\begin{equation}
\label{eq16}
\frac{\partial P_1 }{\partial \xi } = \frac{3\sqrt 3 e^2c\gamma
^4}{2RC}F(\xi ),
\end{equation}

\noindent
where $\beta = v / c$ -- the relative particle velocity, $\gamma =
\varepsilon / mc^2$-- the relative energy, $F(\xi ) = \xi \int\limits_\xi
^\infty {K_{5 / 3} (\xi )d\xi } $, $\xi = \omega / \omega _c $, $\omega _c =
3\beta \gamma ^3c / 2R$ -- the critical radiation frequency, $R$-- the orbit
radius in a bending magnet of the storage ring [6,7]. One can also calculate
$\int_0^\infty {F(\xi )d\xi } = 8\pi / 9\sqrt 3 $[6]. Thus the full
radiation power for one particle can be expressed as:

\begin{equation}
\label{eq17}
P_1 = \frac{4\pi }{3}\frac{e^2c\gamma ^2}{RC}.
\end{equation}

In the case under consideration the radiation is coherent if the wavelength
is longer than the length of the bunch i.e. $\xi < < 1$, $K_{5 / 3} \left(
\xi \right) \approx 2^{4 / 3}\Gamma \left( {2 / 3} \right)\xi ^{ - 5 / 3}$,

\[
\int\limits_\xi ^\infty {K_{5 / 3} (\xi )d\xi } = \int\limits_0^\infty {K_{5
/ 3} (\xi )d\xi } - \int\limits_0^\xi {K_{5 / 3} (\xi )d\xi }\]
\[ = \pi \sqrt 3
- \int\limits_0^\xi {K_{5 / 3} (\xi )d\xi } ,
F(\xi ) = 2^{2 / 3}\Gamma (2 / 3)\xi ^{1 / 3}.
\]

Now the formula (\ref{eq16}) can be written as

\begin{equation}
\label{eq18}
\frac{\partial P_1 }{\partial \xi } = \frac{3\sqrt 3 e^2c\gamma ^4}{2^{4 /
3}\pi R^2(1 + \mu )}\Gamma (\frac{2}{3})\xi ^{1 / 3}.
\end{equation}

The spectral coherence factor $s(\omega )$ is determined by the particle
density distribution law $\rho (z)$ and for the Gaussian distribution

\begin{equation}
\label{eq19}
\rho (z) = \frac{1}{\sqrt {2\pi } \sigma _x }e^{\frac{ - z^2}{2\sigma _x^2
}}
\end{equation}

\noindent
can be derived from equations (\ref{eq13}) and (\ref{eq19}) as
$s(\omega ) = \exp ( - 4\pi ^2\sigma _x^2 / \lambda ^2)$ [8,9]. The
value $\sigma _x $ is the mean square bunch length.

The full power of the spontaneous coherent SR, the average loss rate for a
single particle and the losses over a revolution can be calculated
numerically using the formula (\ref{eq15}) and the expression $P^{coh} = f \cdot
\varepsilon ^{coh}$. In the special case when the coherent SR is dominated
by the low frequency radiation $\xi < < 1$, taking into account (\ref{eq18})
and$\int\limits_0^\infty {k^{1 / 3}\exp ( - k^2\sigma _x^2 )dk = \Gamma (2 /
3) / 2\sigma _x^{4 / 3} } $, one can derive that

\begin {equation}
\label{eq20}
P^{coh} = c\frac{d\varepsilon ^{coh}}{dt} = \frac{3^{1 / 6}\Gamma ^2(2 /
3)r_e cN^2}{2\pi R^{2 / 3}\sigma _x^{4 / 3} (1 + \lambda )}mc^2,
\end {equation}

The energy losses per a revolution are

$$\Delta \varepsilon ^{coh} = \frac {d\varepsilon ^{coh}}{dt}T =
\frac{3^{1 / 6}\Gamma ^2\left( {2 / 3} \right)r_e R^{1 / 3}N^2}{\sigma
_x^{4 / 3} }mc^2 $$

\begin {equation}
\label{eq21}
\approx 3.1 \cdot 10^{ - 7} {R^{1 / 3}N^2}/{\sigma
_x^{4 / 3} }[eV/revolution]. \end {equation}

The formula (21) matches with the results of the first work on the
coherent SR [10], is $2^{1 / 6} \approx 1.12$ times lower than one in the
reference [4] , $2^{7 / 3} \approx 5.04$ times lower than the value in the
reference [11] and $2^{8 / 3} \approx 6.35$ times lower than one in the
reference [12]. In the last reference the authors used formula from the work
of Shiff [10] and erred in converting it to their definition of the
value$\sigma _x $. They multiplied the Shiff's formula by the 2$^{4 / 3}$
instead of dividing by it. In the remaining references the source of errors
is unclear but more probably connected with the same mistake.

The coherence factor is decreasing for the wavelengths $\lambda \ge \lambda
_d = 2\pi \sigma _x $ or if $\omega \ge \omega _d = c / \sigma _x $. The
expression (21) is justified if the main part of the energy of the
coherent SR is emitted in the spectral range $\omega \le \omega _d \approx
\omega _c (\xi < < 1)$ i.e. when $\sigma _x > \lambda _c / 2\pi $,
where $\lambda _c = 2\pi c / \omega _c = 4\pi R / 3\gamma ^3$. The
expression also (21) doesn't take into consideration the shielding of
the beam by the vacuum chamber, which leads to the weaker radiation for
the wavelengths longer than the vacuum chamber gap.

The vast majority of the energy is emitted in the angular range \textit{$\Delta \theta \sim $1/$\gamma $} relative to
the direction of the particle's motion when $k \cdot r\left( {t}{'} \right) <
< k \cdot r / \gamma $. So, the condition when one can neglect the
transversal beam dimensions is $k_d \cdot r < < \gamma $ or $r < <
({\lambda _d }/{2\pi })\gamma = \sigma _x \gamma $.

\textbf{References}

1. A.A. Kolomensky and A.N. Lebedev, Theory of Cyclic Accelerators.
North Holland Publ., , 1966.

2. H. Bruk, Accelerateurs Circulaires de Particules (Press
Universitaires de France, 1966).

3. H. Wiedemann, Particle Accelerator Physics I $\backslash ${\&} II
(Springer-Verlag, New York, 1993).

4. L.V.Iogansen, M.S.Rabinovich, JETP, v.37, 118, 1959 (in Russian).

5. J.B.Murphy, S.Krinsky, R.L.Gluckstern, Particle Accelerators, 1997,
v.57, pp.9-64.

6. D.Ivanenko, A.Sokolov, Classical Theory of field, GITTL, 1951 (in
Russian).

7. L.D.Landau, E.M.Lifshits, The classical Theory of Fields, Pergamon
Press, Oxford, UK, 1975.

8. H.Wiedemann, Particle Accelerator Physics, v.I, Springer -- Verlag,
NY, 1993, p.319.

9. A.Andersson, M.Johnson, B.Nelander, Coherent synchrotron radiation
in the far infrared from 1 mm electron bunch, Opt. Eng. V.39, p.
3099-3115, 2000.

10. L.Schiff, Rev. Sci. Instr., v.17, 7, 1946.

11. E.L.Saldin, E.A.Schneidmiller, M.V.Yurkov, TESLA FEL Report 96-14,
November 1996.

12. J.S.Nodvic, D.S.Saxon, Phys. Rev., v.96, No 1, p. 180-184, 1954.

\end{document}